# Biological Extension of the Action Principle: Endpoint Determination beyond the Quantum Level and the Ultimate Physical Roots of Consciousness

Attila Grandpierre
Konkoly Observatory of the Hungarian Academy of Sciences
H-1525 Budapest, P. O. Box 67, HUNGARY
E-mail: grandp@iif.hu

## Abstract

With the explosive growth of biology, biological data accumulate in an increasing rate. At present, theoretical biology does not have its fundamental principles that could offer biological insight. In this situation, it is advisable for biology to learn from its older brother, physics. The most powerful tool of physics is the action principle, from which all the fundamental laws of physics can be derived in their most elegant form. We show that today's physics is far from utilizing the full potential of the action principle. This circumstance is almost inevitable, since it belongs to the nature of the physical problems that the endpoint of the action principle is fixed already by the initial conditions, and that physical behavior in most cases corresponds to the minimal form of the action principle. Actually, the mathematical form of the action principle allows also endpoints corresponding to the maximum of the action. We show that when we endow the action principle with this overlooked possibility, it gains an enormous additional power, which, perhaps surprisingly, directly corresponds to biological behavior. The biological version of the least action principle is the most action principle. It is characteristically biological to strive to the most action, instead of manifesting inert behavior corresponding to the least action. A fallen body in classical physics cannot select its endpoint. How is it possible that a fallen bird can select the endpoint of its trajectory? We consider how the photon "selects" its endpoint in the classical and the extended double-slit experiments, and propose a new causal interpretation of quantum physics. We show that "spontaneous targeting" observed in living organisms is a direct manifestation of the causally determined quantum processes. For the first time, we formulate here the first principle of biology in a mathematical form and present some of its applications of primary importance. We indicate that the general phenomenon of biological homing relies on long-range cooperative forces between biomolecules, including mechanical, electromagnetic and osmotic forces. We show how theoretical biology beyond the quantum level can shed light to the properties of elementary consciousness.

## Introduction

Natural sciences are divided into two main branches: physics and biology. Nature itself is not divided to physics and biology. Therefore physics and biology must go back to a deeper common basis at the most fundamental level of Nature. It is of primary importance to explore the full potential of most fundamental principles of physics and biology.

It seems that we have entered to a new century of biology, in which "the new frontier is the interface, wherever it remains unexplored" and "progress is based ultimately on unification" (Kafatos and Eisner 2004). Recently, biological physics became a new frontier of natural sciences (Phillips and Quake 2006; Sung, 2006). Biological physics is the interdisciplinary effort to cross the barriers between physics and biology from the biology side (Sung, 2006). While biophysics considers the biology-physics interface from the physics side, biological physics seems to be predestined to embrace physics on a biological basis.

We have an enormous amount of data in biology. But we do not have a general theory. The US National Science Foundation allocates billions of dollars year by year for initiating the birth of theoretical biological physics (Ladik, 2004). "Today, by contrast with descriptions of the physical world, the understanding of biological systems is most often represented by natural-language stories codified in natural-language papers and textbooks…But insofar as biologists wish to attain deeper understanding (for example, to predict the quantitative behavior of biological systems), they will need to produce biological knowledge" (Brent and Bruck, 2006). In the absence of real prospects for obtaining deeper insights, a growing frustration can be observed in the community of biologists. In a sense, they are right. Biology's right to exist as a science asks for a theoretical background and the formulation of its fundamental principles.

It is clear that the usual bottom-up approach of physics cannot be successful in biology because of the overwhelming complexity of living organisms. Therefore, theoretical biological physics puts into the focus the effort to develop new principles and models for complex systems based on biological phenomena. In this paper we consider here for the first time the interface between physics and biology in its most fundamental aspects, i.e., at the level of first principles, presenting the mathematical, physical and biological background of the first principle of biological physics, indicating some of its applications as well. What we need is the minimal extension of physics into biology. The aim is to learn from physics as much as possible, and expand it as little as necessary to make it suitable to describe biological behavior. Both of these aims are possible to realize by only one „small" step: using the action principle *with endpoint selection.* Endpoint selection is an objective, natural fact of primary importance for biology. Its consequences are physical and biological; measurable, objective facts. The living bird's trajectory differs from that of a dead bird in an objective, measurable manner.

"Biology, the scientific study of living organisms, is concerned with both mechanistic explanations and with the study of function." (Purves et al., 1992) The fact that biology must present not only the mechanistic, but also functional explanations, is directly related to the fact that while in classical physics the final state is determined by the initial conditions plus the physical laws, in biological organisms the case is the opposite. We argue that this property, the surplus of functions over mechanistic terms so fundamental in biology, is intimately related to the maximum version of the action principle, and to its primary property: endpoint selection.

Obtaining quantitative laws of biological behavior at the level of organisms by physical methods seems to be impossible by reason of the intractable complexity of the organisms themselves. We derived a method that starts, not from the level of material constituents, but from the level of behavior as determined by means of an extended version of the physical action principle.

**Endpoint determination and the full potential of the action principle**

The action principle is widely acknowledged as the first principle of physics. Yet the term "first principle" is somewhat vague and requires a precise definition. We consider a branch of natural sciences as "mature" if and only if its "first principle" is formulated in a mathematical form that can be widely used to derive fundamental equations describing actual behavior. We propose to call a fundamental law of physics by the term "first principle of physics" if and only if all of the fundamental laws of physics — e.g., that of classical mechanics, hydrodynamics, electromagnetism, thermodynamics, theory of gravitation, and quantum physics, including quantum field theories and string theory — can be derived from it. This means that the action principle is the most powerful tool of physics, having a capacity at least that of the fundamental equations all together. Actually, it is even more powerful, since its integral aspects present an additional and fundamental potential. Similarly, a first principle of biology, if it exists, has to be one from which all the fundamental quantitative laws of biology can be derived. This means that the first principle of biology is the most powerful tool of theoretical biology. The solution we have found represents an important step towards establishing the first principle of biology and contributes to an opening up of a new frontier of science that is capable of an unexpectedly elegant unification of physics and biology.

Action is the tool of choice when we want to specify constraints on both initial and final conditions. A fallen body in classical physics cannot select its endpoint, since it is already determined at the outset. How is it possible that a fallen bird can select the endpoint of its trajectory? In classical physics, when we specify the initial state, this automatically specifies the end state through the deterministic equations. When we drop a ball from a certain height, its end state is already determined physically. The case must be different in biology; for in living organisms, the end states must be consistent with biological functions like survival. The endpoint of a living bird's trajectory dropped from a height cannot be the trajectory endpoint realized by a ball or a dead bird. In physical systems the final state is determined by the initial conditions plus the physical laws. In biological organisms the case is the opposite: the final state is determined by the living organisms themselves (Bertalanffy, 1950). Action is an ideal tool of choice to describe biological phenomena since it can act also on biologically determined endpoints. Today's physics does not utilize the full complement of the action principle; namely, endpoint determination, maximum form, and minimum form. In physics, the end points are fixed, and the action is in most cases a minimum (Brown, 2005, xiv). Yet the enormous potential of the action potential in explaining Nature enfolds in its real power only when we set free the action principle from these constraints, and allow the selection of the endpoint and of the maximal action as well. Liberating the action principle from its straitjacket one can expect to obtain a deeper insight and a broader picture of the most fundamental aspects of Nature.

Action principles can seem puzzling to the student of physics because of their seemingly teleological quality: instead of predicting the future from initial conditions, one starts with a combination of initial conditions and final conditions and then finds the path in between, as if the system somehow knows where it's going to go. The path integral approach (Feynman and Hibbs 1965) is one way of understanding why this works. The system does not have to know in advance where it is going; the path integral sums over the probability amplitudes for all possible processes, and the stationary points of the action mark neighborhoods of the space of histories for which quantum-mechanical interference will yield large probabilities. We point out that the quantum mechanical interference in a cosmic context relies on the existence of virtual processes occurring instantaneously, mapping the whole of the universe before the start of the quanta. Teleology can be exiled only at the cost of instantaneous "mapping" interactions of the whole situation globally, plus allowing the ability of quanta integrating all the probabilities of all possible paths, possibly involving also the cosmic context. This integrating ability arises from the integral aspects of the action principle, and it is an addition to the differential equations frequently associated with local causality in the mechanistic approach.

The teleology of freely selected endpoints presents an even harder problem for the pan-mechanistic hypotheses. Some natural systems seem to be able to select endpoints or final states – these are the living organisms. The only possibility for such selection offered within the conceptual frameworks of physics is given by introducing random probabilities. Nevertheless, non-randomly selected endpoints transcend the conceptual framework of physics when it corresponds to lawful behavior, like in biology where they arise. The local causality viewpoint relying on the differential equation formalism ignores the integral aspects of the action principle, and so rejects to form a more complete picture of Nature. As Stapp (2003, 53) pointed out, the claim that "goal-directed activity can always be reformulated casually [by solving differential equations]" is not applicable to the process of endpoint selection in biology, even if it is applicable after the endpoint is selected. The process of endpoint selection is additional to physics working only with a part of the potential of the action principle, without endpoint selection.

It is easy to realize that teleological concepts and explanations are not inherently obscure (Nagel, 1979, 314). Instead, they are quite compatible with biology, even if they are not compatible with pan-mechanistic assumptions. "The causal explanation cannot rightly be charged with the difficulty often raised against teleological explanations, that the causal explanation assumes that a future state of affairs can be causally efficacious in bringing about its own realization. For according to the correct explanation, it is not the goal, itself existing in the future, that brings about the action. Instead, it is rather the agent's wanting the goal, together with his belief that the action would contribute to the realization of the goal that does so" (Nagel, 1979, 278). Explanations of goal ascriptions in biology are therefore causal (ibid., 314).

It seems that the best interpretation of the action principle of physics is the path-integral method that works at the level of quanta and virtual interactions. Feynman's path integral approach indicates that *quanta explore all possible paths between the initial and end states* (Taylor, 2003; Moore, 2004), and the resulting path is the integrated sum of all these paths. The path integral interpretation of Feynman is fitted as much as possible to the presently dominant mechanistic scheme, apparently ruling out the need to allow a kind of consciousness to be present at the level of quanta which could pre-select the optimal path. Yet we point out that the exploration of all paths possible in the actual situation – the phenomenon of "quantum orientation" - seems to transcend the mechanistic scheme, in the sense that it would require non-local, instantaneous connections without entanglement. Moreover, the integration of the probability amplitudes of all possible paths also cannot be described in the differential equation formalism, therefore, it transcends the local causality scheme of mechanistic philosophy. The ability of quanta to orientate themselves with the help of virtual (note: the word has the connotation 'non-physical') interactions do not fit well to the pan-mechanistic hypotheses. Moreover, the integrative ability of quanta points to the presence of an integral aspect, and this integral aspect, as we indicated above, transcends the pan-mechanistic scheme. Actually, this central problem of quantum physics is still not solved (see below). Indeed, other interpretations seem equally possible. For example, Zukav (1980, in the chapter "Living?", pp. 45-66) argues that "Something is "organic" if it has the ability to process information and to act accordingly. We have little choice but to acknowledge that photons…do appear to process information [in the two-slit experiment] and to act accordingly, and that therefore, strange as it may sound, they seem to be organic" (ibid., pp. 63-64).

In order to explore how the living bird can select its endpoint, we look for the apparently simpler problem how the photon "selects" its endpoint in the crucial experiment of quantum physics, the double-slit experiment.

## Quantum orientation

Quantum experiments such as the classic double-slit experiment showed that the single photon somehow "knows" whether the two slits are open or not, even if it goes through only one of them as a particle. It was Niels Bohr who pointed out in his celebrated paper "Light and Life" (Bohr, 1933) that the quantum nature of light has profound implications for biology. Yet the spatial continuity of light propagation on the one hand, and the particle nature of the light effects on the other seemed to present a conceptual difficulty. He concluded that "This very situation forces us to renounce a complete causal description of the phenomena of light and to be content with probability calculations."

Since Bohr made this claim, we have learned that the interpretation of quantum mechanics is not a settled question. There are many different interpretations, and there seems to be no consensus about which one is correct. We think it is not without reason that "nobody understands quantum mechanics" (Feynman, 1990, 9), keeping in mind Einstein's famous claim, "If it [the Copenhagen interpretation of quantum mechanics] is correct, it signifies the end of physics as a science." Many outstanding physicists expressed similar opinions (including Bohr, Schrödinger, Wheeler, etc.). By our evaluation, the general confusion is rooted in an attitude that seems to reject the principle of causality itself.

Therefore, we propose a new interpretation of quantum physics that would allow for the first time to keep the causal description as consistent with all facts, yet would enable us to connect quantum physics with biology in an unexpectedly simple and elegant way. Such an interpretation offers a glimpse into the realm beyond the present-day quantum physics. It is well known that all fundamental quantum physics, according to Feynman (1967, 130), can be illuminated by comparison to the double-slit experiment. "In reality, it [the double slit experiment] contains the only mystery" [of quantum physics] (Feynman, Leighton and Sands, 1965, Vol. 3, 1). It is a growing understanding that chemical "double slits" (Dixon et al., 1999; Pophristic and Goodman, 2001; Weinhold, 2001) and multi-slits (Hackermueller et al., 2003) may play a central role in chemistry and biology.

We point out that the causal interpretation of quantum physics we present here make it possible for the first time to solve this central problem of quantum physics. Presenting a natural extension of this causal quantum interpretation into biology we obtain an underlying basis for the working mechanism of the action principle in biology.

Formulating the biologically extended version of the action principle it becomes possible to test empirically the causal interpretation of quantum physics in the case of living organisms, therefore deciding in favor of mechanistic or biological interpretation of the double slit experiment.

The plausible extension of the causal quantum hypotheses into biology involves only a simple and natural step. We propose that not only in the double-slit experiments but also in more complex physical and biological systems, somehow *a photon, through its wave nature, perceives the whole manifold of possible paths immediately, and realizes that path which corresponds to the actual physical and biological situation as a whole.* This proposal expresses *a biological interpretation of quantum orientation within living organisms.* We do not claim only that the photon perceives the living organism as a gigantically complex system of coupled multi-slits, but also that it observes the "biological situation" as well; i.e., the photon explores all details of all paths, including the most fundamental biological coupling of the global thermodynamic parameters of the organism that is the maintenance of the distance from equilibrium, which we will consider below in more details. Therefore we offer a definite and plausible physical basis for the term "biological situation" when suggesting a biological interpretation of the key experiment of quantum physics.

Recent evidences seem to underpin the biological role of the wave functions in living organisms. The two-dimensional Fourier transform electronic spectroscopy of photosynthetic complexes has mapped the excited energy levels and clearly document the dependence of the dominant energy transport pathways on the spatial properties of the excited-state wavefunctions of the whole bacteriochlorophyll complex. This wavelike characteristic of the energy transfer allows the complexes to sample vast areas of phase space to find the most efficient path (Engel et al., 2007). Sension (2007) notes that the observed 'quantum beats', which persists for hundreds of femtoseconds, are characteristic of coherent coupling between different electronic states. In other words, the electronic excitation that transfers the energy downhill does not simply hop incoherently from state to state, but samples two or more states simultaneously, securing the exquisitely tuning to capture solar light efficiently.

Among others these empirical results present serious indications in favor of the role wavefunctions in facilitating biological functions, underpinning the here proposed biological interpretation of quantum orientation. Now we think that if biological organization is capable to harness the virtual interactions that are the underlying base of the action principle, than biology must act at a deeper than quantum level.

**On the physical meaning of action, the central concept of physics**

It is a truism that the *physical meaning* of each symbol contained in any principles of physics has to be specified before the theory can be applied in practice (Yourgrau, Mandelstam, 1955, 139). The action principle of physics is claimed to be, regarding its form and content, to come nearest to the ideal final aim of theoretical research to condense all natural phenomena into one simple principle, that allows the computation of past and future processes (ibid., 126). Actually, it seems that the physical meaning of what action is, remains obscure. The saying "I don't know what action is" is also attributed to Feynman (Toffoli, 2003). Similarly, in the Encyclopedia Britannica (1988, 1: 71) it stands: "action is an abstract quantity that describes the overall motion of a physical system", and this indicates that that the physical meaning is not explicit. It is not without reason that action shows up in the "Glossary of Frequently Misused or Misunderstood Physics Terms and Concepts" (http://www.lhup.edu/~dsimanek/glossary.htm). Pascual Jordan remarks that "Although the concept of action is less obvious to man's physical intuition than that of energy, it is of even greater significance, as it appears also in connection with the quantum laws" (Jordan, 1988, 25: 681).

Yourgrau and Mandelstam also acknowledge that "With the development of the older form of quantum theory the persuasion that the action had some deeper meaning gained renewed impetus" (ibid., p. 128). Nevertheless, their result seems to point towards only a mathematical meaning and excluding any possible physical meaning, as it is shown by Yourgrau and Mandelstam definite claim: "The action function can be fully and satisfactorily defined in terms of the other constructs and laws of dynamics, and it is thus rather an invaluable mathematical aid than a means of interpretation" (ibid., 140).

In this way, we find that the situation at the very foundations of physics is paradoxical. Although its very central concept, action, is formulated mathematically, its physical meaning is obscure. At the same time, physical meaning has central importance in the progress of physics. In the absence of a closer understanding of the physical meaning of action the progress of physics seem to be restricted to stagnation in respect to one of its most fundamental aspects. The action is the "number of related integral quantities which serve as the basis for general formulations of the dynamics of both classical and quantum-mechanical systems" (McGraw-Hill Science and Technology Encyclopedia, 2007). The most fundamental quantity of the action principle, the action is a cost function (Rosen, 1967, 4, 155). This special cost function measures the product of energy investment and time investment. On the basis of the pan-mechanistic hypotheses, it would be plausible to consider that time investment cannot be a factor determining the behavior of an inanimate system. It is clear that energy investment and time investment, and, especially their integrated product function, measures a quantity having a biological aspect. The appearance of the biological aspect at the ultimate level of physics may seem as paradoxical. We will refer to this paradox as *the biological paradox of physics*.

We can add to this biological paradox *the teleological paradox* that also entered into physics at its ultimate base with the action principle. We found also that the most crucial experiment of quantum physics, the double slit experiment is related to problems transcending the conceptual framework of physics, because quantum orientation, integration and biologically useful functioning all indicate that these fundamental problems of quantum physics can be solved only at a deeper, biological level. It seems that a biological interpretation of quantum physics is inevitable. This is the third biological paradox found at the most fundamental level of physics: the biological interpretation paradox.

For us, it seems that these paradoxes together can be regarded as more than remarkable, when thinking about the next frontier of physics. Actually, the existence of a biological principle throughout the Universe, at the level of elementary interactions as well as at the cosmic level is already recognized in the "*principle of elementary interactive perception*" (Grandpierre, 2000).

## The principle of interactive perception of quanta

Endre K. Grandpierre had pointed out (Grandpierre, 2000) that the concept of interaction already assumes a dynamic unit of action and reaction. In an elementary interaction, action and reaction form an instantaneous unit, implicitly involving a kind of elementary circularity: the re-action instantaneously re-acts to the action, forming a closure of the elementary interaction. Moreover, the reaction of the 'second' object also elicits a reaction, namely, a reaction of the 'first' object to the reaction, and so on, forming a recursive scheme that constitutes a complete elementary event: the elementary interaction. Definitely, the escalating chain of actions and recursive reactions must form a 'final' closure in order to obtain a definite event. This means that the dynamism of action and reaction fundamentally involves a kind of *"direct" circularity,* a kind of other-reference as well as self-reference. The reaction to the acting entity carries always the traces of the action, and so the consecutive series of reactions represent a recursive series of reflected actions. It is apparent that this kind of other-reference and self-reference is what actually governs the arising behavior of the acting entity as a whole, at its global level, and so we reached to an elementary form of the most fundamental properties of consciousness: other-reference and self-reference.

Exploring the directly circular logic of the elementary interaction is not the end of weird phenomena found at the microworld. This fundamental "direct circular" complexity of the most elementary event of interaction is accompanied by an additional *"laterally circular"* complexity, because all the virtual particles involved in establishing the interactions are themselves possible objects of further virtual interactions, as represented by higher and higher order side-loops in the Feynman diagrams (Feynman, 1990, Figs. 73-78, pp. 115-127). These sideward "corrections" also represent an infinite sequence. As quantum electrodynamics teaches us, the exchange of virtual quanta involves side-branches, additional loops of virtual interactions, in an infinite cascade. There will be an infinite number of possible interactions all going on at once already within one single interaction. But even a single electron interacts through its electric and gravitational charges with the whole of the material universe. Therefore, we also must envisage the *astronomically multiple interactions* between all particles as produced by webs of interacting complexity made up of ever more convoluted exchanges between different sorts of virtual particles. In a practical calculation of mechanical forces between particles it is unusual to consider more than three or four of the simplest diagrams unless very high accuracy is required (Davies, 1984, 105). Importantly, we have to point out that the case is different when we consider situations in which complexity is relevant. Moreover, the viewpoint of practical calculations is fundamentally different than the viewpoint of theoretical interest. For example, when we consider loops of interactions between elementary interactions of an astronomical number of contributors, like in biology and, definitely, in the cosmic context we envisage here, too, the progressively smaller contributions of higher order "corrections" can add up to significant amplitudes. These significant amplitudes may form the natural basis of collective interactions characteristics in biology, and, possibly, in cosmology.

Regarding the problematic aspects of interactions, we are inclined to give up the picture in which the material objects are the fundamental entities, and, instead, find a deeper basis of existence in interactions. This argument is well underpinned by the fact that in quantum physics the observable quantities are interactional properties. In quantum physics, "the observed object and the observing system are involved in an irreducible and unanalysable interaction. Since the results of observation derive from this interaction, they must be taken to represent properties of the interaction as a whole, not properties of the observed object alone" (Villars, 1984). The primacy of interactions over objects has profound consequences, as it is indicated also by e.g. the delayed choice experiment. And because the interactions form an interwoven network, through direct and lateral circularity and the astronomically multiple nature of interactions, we arrived to the indication of a *fundamental non-metricity* at the level of interactions. All interactions together form one single unit of an extremely sophisticated cosmic network. In this basic fact we recognize the primary interconnectedness of interactions, and, through them, the primary interconnectedness of the material universe. In other words, when we refer to the whole system of interconnected network of interactions as the vacuum, we can say that the vacuum does not exist within the three dimensional space of physical observables. Indeed, in quantum electrodynamics, the vacuum is the source of virtual and manifestly observable interactions, too.

Obtaining a preliminary introduction into the nature of the principle of interactive perception of quanta (Grandpierre, 2000), we have to add that in this picture the most elementary form of interaction as a dynamic unit of directly circular and sideward circular infinitely complex processes represents an elementary form of perception, corresponding to the "action" and "reaction" aspects of this complex phenomenon that is in some sense paralleled with the most elementary form of "stimulus" and "response". In this way, already the most elementary interaction corresponds to an elementary form of perception. Without the ability to form a dynamic unit from the infinite chains of directly and laterally circular and astronomically multiple interactions, the whole universe would disintegrate and become unable to exist. Elementary perception is the ultimate basis of material existence.

Grandpierre (2000) also pointed out that elementary interactions are intimately related to energy. Regarding the aspect of energetics, the collective nature of interactions involves that the energy of the elementary interactions demonstrates a par excellence collectivity. Energy is dynamic, oscillating, collective, and directly unobservable; therefore, if we define matter as the entity that is directly observable through our outer senses, than energy is an immaterial form of existence. In contrast, matter is inert, rigid, passive, individual, and inanimate. Energy is a circularly coupled form of the dynamic process of the action-reaction event. The energetic interactions of the universe transform the static concepts of existence onto a dynamic footing, where the fundamental entities of the universe are events having a cascading, interactive, and extending nature. Such energetic interactions couple the whole of the universe into a dynamic process. The population of the universe with such fundamentally dynamic energetic interactions in cascaded hierarchies shows a kind of similarity with the fecundity principle of life. The whole universe appears as a gigantic and throbbing thread of inevitably propagating and complexifying chain reaction of interactions, including all the known and yet unknown forms of interactions, elevating the universe to higher and higher levels of organization, creating spontaneously self-active systems of activity. In this way, the principle of interactive perception becomes the basis of the upward organization of the whole universe. *We propose that the quantum orientation observed in the two-slit experiment is a direct manifestation of the perceptive interaction of quanta.*

### Spontaneous organization: spontaneous targeting and biological homing

Indeed, we observe in Nature that all objects with free energy and movable surfaces spontaneously organize themselves into systems like atoms, molecules, and more complex systems. Many aspects of this *spontaneous organization* are automatic and universal; therefore we usually ignore them. Actually, spontaneity is a remarkable property, especially when a compound system builds up from its elements. For example, it is generally thought that water molecules form automatically and spontaneously, if the mix of relevant elements is supplied with the necessary amount of energy to trigger the process. But the question arises: how do the molecules of hydrogen and oxygen find each other and meet with just the right energies, angles, and distances that are the necessary conditions of bonding, all accomplished within a fraction of a second? If they follow a Brownian random motion, only a minute part of the molecules with proper impact factors (distance, angle, and velocity) could collide in the manner necessary to form the compound molecules given the explosively fast timescales.

We add that independent evidence underlies the above argument. It appears for us that spontaneous targeting is at work in the phenomenon of the spontaneous formation of water molecules (Hemley, 1995). This reaction between hydrogen and oxygen is surely one of the best studied chemical reactions, but details of the reaction kinetics have always been enigmatic. Loubeyre and LeToullec (1995) found that the normally explosive reaction is shut down by pressure, and a new compound is formed. This means that the factors spontaneously organizing the reactions are sensitive to such *integral* parameters as global pressure and temperature.

The *principle of elementary interactive perception* is widely referred to as "spontaneous targeting" (Grandpierre, 1997; Torchilin, 2007); as "active navigation" (Mikhailov, Hess 1995); as well as "spontaneous insertion mechanism" (Di Cola et al., 2005). Active or spontaneous navigation is indicated to be able to utilize thermal fluctuations through thermal ratchets (Mikhailov, Hess 1995).

In biological thermodynamics, spontaneity is defined by the Gibbs free energy. A process will occur spontaneously only if the corresponding change of the Gibbs free energy is negative (Haynie, 2001, 81). Such processes are termed as exergonic. If our proposal is valid, *a photon is spontaneously emitted and absorbed in a living organism "in due time and place."* Timing is of key importance in biology, since it is the determination of which processes occur where and when; and this is what constitutes the interface with biological regulation. Virtual particles do not interact with any real particles during their lifetime, except at their emission and absorption. This makes virtual particles an ideal tool for biological timing. Recent evidence in budding and fission yeasts indicates the existence of a timing mechanism that is independent of DNA replication (Weinert, 2007). The outcome of the absorption of the photon contributes to the actualization of a biologically needed process. In the biological interpretation of quantum orientation, the newly born photon immediately perceives the internal states of the whole organism, and the photon's path endpoint is directed to that part of the organism that is most in need of preserving the distance from equilibrium, keeping this distance as large as possible. There the photon will be absorbed. *Thus the photon has a critical biological function, as Bohr (1933) suggested.* We point out that the capability of the photon to perceive the situation, and thus become suitable to serve biological needs, could be a key to the development of a theoretical biology that can reach one level deeper into nature's secrets than quantum physics itself.

A spontaneous targeting mechanism seems to work in the olfactory field, as well. How do the odorant molecules automatically find their receptors? Lewis Thomas (1974) noted that eels have been taught to smell two or three molecules of phenylethyl alcohol. An average man can detect just a few molecules of butyl mercaptan. Recently, Brookes et al. (2007) confirmed that electron tunneling mediates the spontaneous targeting of odor molecules. Fagas et al. (2006) by the explicit treatment of electron-electron interactions had shown that such tunneling is the result of many-body quantum effects that *maximize* the overlap of single-particle states. The substance-receptor system may somehow communicate from a distance through a non-local process, resulting in the spontaneous targeting of the substance to the receptor.

Aromacity, covalent and coordinate bonding, resonance delocalization, H-bonding and hyperconjugation all can be seen as special cases of a central superposition paradigm (Weinhold, 1999). Spontaneous donor-acceptor interactions are the manifestations of virtual quantum effects and play a central role in chemistry. The subtle virtual effects modify the electron distributions of the donor and acceptor, and the modified electron distributions determine the behavior of the molecules. Frequently, quantum effects generate optimal overlap between electron orbitals (Pophristic and Goodman 2001). Hyperconjugation, a quantum effect involving the transfer of electrons from an occupied orbital to an unoccupied orbital, plays also an important role in determining the potential barriers.

It is already noticed that such spontaneous targeting processes play a crucial role in biology. Meggs (1998) realized that the sophisticated orchestration of biological forces, each exerted in due time and due place, is something that seems to be a miracle from the viewpoint of physics. He illustrated it by protein synthesis. "A double stranded DNA molecule unzips, and an enzyme called RNA polymerase migrates to the site and attaches itself to one of the strands. The RNA polymerase then ratchets along the DNA. At each step, a specific ribonucleotide migrates to the site and attaches to a growing chain of such molecules to form an RNA molecule. The resulting RNA then snakes out of the cell nucleus and moves to a structure in the cytoplasm called a ribosome. The RNA threads itself through the ribosome in discrete steps, and at each step a specific amino acid migrates to the ribosome and attaches to a growing chain of amino acids to form a protein. A physicist, however, remains befuddled. *What forces move just the right ribonucleotide molecule into position to form the RNA chain just the right time and place? What forces move the RNA molecule from the nucleus to the ribosome, and what forces bring the right amino acid to the ribosome at each step in forming the protein chain?*

Meggs (1998) introduced the term "biological homing", referring to such spontaneous targeting mechanisms: Hormones bind receptors, antibodies targets to antigens, enzymes to their substrates, actin molecules combine with great specificity and rapidity with actin filaments, tubulins with microtubules, receptors with neurotransmitters. Biological organization consists in generating the specific factors that activate or transport the right biomolecules at the right place at the right time. Meggs (1998) argued that *the homing force has an electromagnetic nature* and arises from complementary distributions of electric charges on the surface of molecules. He has shown how the enhanced attraction of complementary pairs of molecules can be calculated using quantum mechanics. Certainly, a long-range interaction cohering interactions with the functions of the cell as an additional mechanism must also exist that generates the complementary distributions of electric charges, based on a highly specific interaction between the targeting molecules and their targets. We suggest these fundamental mechanisms are related to quantum orientation we considered above.

These observations make it clear that a yet largely unnoticed *spontaneity* acts as an organizing factor in the quantum world, alongside and beyond the already considered phenomena of physical self-organization, based on quantum orientation and non-local, virtual interactions.

By our proposal, these many-body effects are based on quantum orientation, which relay on virtual interactions, and correspond to an elementary form of consciousness. In this way, the above arguments all indicate that consciousness (perhaps a better term would be proto-consciousness) must be present at the most fundamental levels of matter and universal vacuum fields.

## Consciousness and quantum orientation

It is a widely acknowledged view in biosemiotics that life and consciousness are coextensive (Hoffmeyer, 1996, 2001). Therefore, we can also formulate the above indicated (proto)biological interpretation of the two-slit experiment in the following way. The elementary quanta of physics are coupled to the vacuum and manifest an elementary or proto-consciousness. The means of proto-communication are the virtual interactions. The presence of the *consciousness aspect* is one reason to regard these virtual interactions as transcending physics and corresponding to biology. Moreover, these virtual interactions are immediate, representing instantaneous interactions (not necessarily quantum entanglement). This is another reason to regard virtual interactions as proto-communication, as expressions of proto-consciousness.

*We found the following properties of proto-consciousness: perceptive interactions, self-referential activity, quantum orientation, spontaneous timing, spontaneous targeting, and spontaneous upward organization.* Szent-Györgyi (1968, 1972) pointed out that spontaneous electron transfer may be one of the most common and fundamental biological processes. In this way, a natural connection is found between the behavior of elementary particles and biological organization. These findings are in direct relation with the working mechanism of the action principle and its path-integral formulation, with the conceptual and quantum electrodynamic analysis of interactions, with already observed biological phenomena like spontaneous targeting etc. All these factors together make a difference from the somewhat similar approaches of particle consciousness, from Empedocles (5[th] century B. C.), Bergson (1907); to some aspects of process metaphysics (Whitehead, 1929); protomental properties (Nagel, 1979), protobiology (Matsuno, 1989), protophenomenal properties (Chalmers, 1996), panenexperiantialist physicalism (Griffin, 1997), panexperiantialism (Rosenberg, 2004), panpsychism (Seager and Allen-Hermanson, 2005), quantum interactive dualism (Stapp, 2007); and consciousness as the fundamental property of matter (C-property, Scaruffi, 2006).

We consider that spontaneous targeting is just a sub-phenomenon of the more general spontaneous phenomena of *upward organization* (Grandpierre, 2000), which extends from virtual interactions to the Universe as a whole. In this paper, we present a direct relation of this "upward organization" to the biological (maximum) version of the action principle, already established in physics. With the help of a causal interpretation of quantum physics, we indicate that this tendency is rooted in the nature of elementary particles and their couplings to the quantum vacuum, and, ultimately, in the fundamental principle of perceptive interactions.

In our picture of biology, the "behavior" of the photon exemplifies an extended version of the action principle, which in the case of living organism grabs the whole physical plus biological situation, mapping the coupled system of biological energy states. Now if the nature of the photon is connected to the action principle, and biological processes rely largely on photons, then biology has to be fundamentally connected to the action principle. This circumstance offers the conjecture that biology must work on the basis of the action principle through the mediation of light (and, in general, of quanta) and *quantum orientation*. Quantum orientation is intimately connected to the *integral* character of the action principle. In physics, the integral character of the action principle is often ignored. Rather, it is commonplace to regard the differential equations of physics as being "equivalent" with the integral action principle. Definitely, thermal and statistical processes remain present within living organisms; but in biology, the integral character of the action principle acquires *central* footing. And this makes a big difference.

## Photons and biological coupling

Indeed, Bohr (1933) noted that from a physical point of view, light may be defined as the transmission of energy between material bodies at a distance. Life in its physical fundament is energy transmission between the excited states of complex systems. We propose that biological organization acts through couplings such as those between endergonic and exergonic biochemical processes, and the best candidate for the facilitation of such couplings are photons, since they are created in spontaneous processes that are out of the reach of complete physical determination regarding their initial position and timing (spontaneous emission) and final position (spontaneous absorption). In the absence of highly specialized couplings, only exergonic emissions and absorptions are allowed. In the case of biological couplings, it suffices if the processes are exergonic at the global level of the organism. This means that *the presence of biological couplings generates additional possibilities* beyond those that would obtain on the basis of the physico-chemical laws alone, *making possible an astronomically large number of endergonic biochemical reactions through coupling them to exergonic processes*.

Bohr and many other physicists have suggested that the reconciliation of the spatial continuity of light propagation with the orientation capability of quanta can be made only through probabilities: "This very situation forces us to be content with probability

calculations…the electromagnetic description of energy transfer by light remains valid in the statistical sense." Instead *we propose to keep both causality and quantum orientation and consider a plausible possibility that quanta behave statistically only in* unorganized *systems*.

Our argument is the following. In thermodynamic systems, all subsystems of the system can be regarded as independent (Landau, Lifshitz 1959). Now if our causal interpretation of quantum physics is correct, then quanta such as photons in such physical systems immediately behave knowing about all these subsystems' statistically independent and random behavior. Therefore they will behave statistically and randomly; and so probability distribution will enter the game. If so, the situation must be fundamentally different in biological systems. As Bohr (1933) also noted, "An understanding of the essential characteristics of living beings must be sought, no doubt, in the peculiar organization, in which features that may be analogous by the usual [classical] mechanics are interwoven with typically atomic [quantum] traits in a manner having no counterpart in inorganic matter… Owing to the very limits imposed by the properties of light, no instrument is imaginable which is more efficient for its purpose than the eye… This ideal refinement suggests that other organs also…will exhibit a similar adaptation to their purpose." This conclusion seems to be confirmed by recent research into e.g., cochlea, which demonstrated the simultaneous optimization of a number of particular functions and structures "at the natural limits of biological possibilities" (Mammano, Nobili 1993). Bohr added that "In every experiment on living organisms, there must remain an uncertainty as regards the physical conditions to which they are subjected, and the idea suggests itself that the minimal freedom we must allow the organism in this respect is just large enough to permit it, so to say, to hide its ultimate secrets from us."

Our findings underpin Bohr's argument and complement it. From another standpoint, this means that the minimal rate of quantum freedom in complex organized systems such as living organisms can sum up to macroscopic levels that are able to offer physically not completely determined possibilities for biological couplings so that spontaneous biological organization can develop. It is worth keeping in mind that in one droplet of water there are as many molecules as stars in the observable universe. Certainly, the microscopic spontaneous processes of quantum nature usually are not able to sum up into a coherent global organization that consequently conveys the governance of processes from physics into the hands of the biological principle. Certain threshold conditions such as a critical rate and quality of complexity must be met. Yet, these are the very threshold conditions of life.

In this way, we can remove an obstacle posited by the "bottom-up" physical methods that require us to regard the entire universe and living organisms as being deducible from ultimate building blocks plus the fundamental laws of physics. Beyond the pair-interactions, many-body interactions are indicated to play a biological role, endowed with the ability of endpoint selection, following the most action principle. Indeed, many-body interactions make a system more than the mere sum of its constituents.

Actually, recent numerical calculations indicated that polarization and charge transfer are based on many-body quantum effects (York & Yang, 1996). McKenzie (2007) emphasized that quantum physics is responsible for the functionality of photoactive biomolecules. Phillips and Quake (2006) argued that in many instances, the machines of the cell are integrated into collections of many parts, often with proteins, nucleic acids, lipids and other molecules working in concert, comprising many interacting degrees of freedom, collective excitations like phonons and magnons. Many-body quantum effects were considered as responsible for many fundamental aspects of cellular organization (Ursell et al., 2007; Garcia et al., 2007; Ababou et al., 2007 etc.).

By this reasoning, we propose to shed more light on the issue with the help of a theoretical biology that can become the next frontier of science beyond quantum physics.

## Biology as more fundamental than quantum physics

Bohr (1933, 458) himself expressed a similar opinion more than seventy years ago: "The asserted impossibility of a physical or chemical explanation of the function peculiar to life would in this sense be analogous to the insufficiency of the mechanical analysis for the understanding of the stability of atoms." In other words: The theoretical biology which has to be born is in a similar relation to quantum physics as quantum physics is to classical physics. Since quantum physics represents a deeper level of reality than classical physics, the conjecture inevitably arises that biology in its mature form must be able to represent a still deeper level of reality than quantum physics. This means that by extending quantum physics one step deeper into the secrets of nature, a new theoretical biology can begin to emerge.

Similar perceptions of this deeper-than-quantum level of reality have already been expressed. For example, theoretical biologist Ludwig von Bertalanffy (1952, 201) indicated his hope that "the attempt will be made to extend principles like that of least action." Eugene Wigner (1969) came up with the idea that biology is a more general science that includes in itself physics as a special subclass: "Since it is rather clear, in retrospect, that physics in the past always dealt with situations which turned out later to have been limited cases… It may well be suggested, therefore, that present-day physics represents also a limiting case — valid for inanimate objects." Wigner (1970) considered inanimate matter "as a limiting case in which the phenomena of life and consciousness play as little a role as the nongravitational forces play in planetary motion."

## Quantitative theoretical biology from the action principle

But how to utilize Bohr's insights in the construction of a new biology that can reach one level deeper into reality than quantum physics? We propose to start with the action principle, so to construct the Lagrangian for living organisms. We will see that we do not have to do now more than that, since the connection with the usual physical action principle will arise naturally from the problem itself.

It is well known that for mass points moving in a potential field, the Lagrangian is the difference of the kinetic and potential energies. Nevertheless, in living organisms the main type of energy that corresponds to biological functions is *free* energy. More exactly, we can introduce a precise measure for the entropic distance from thermodynamic equilibrium, what is known as *ex*tropy of the nonequilibrium system, $\Pi$ (see Martinás, 1998; Martinás, Frankowicz 2000). The energy of a living organism above its equilibrium level having a temperature $T_0$ will be the *ex*tropic energy $E = \Pi(\text{system}) * T_0$. The *ex*tropy $\Pi$ can be calculated as $\Pi = \int \Sigma X_i dx_i$, where the $X_i$'s are the extensive variables, while the $x_i$'s are the intensive variables in the form of generalized thermodynamic forces. In simplified mathematical terms, $\Pi = U(1/T_0 - 1/T) + V(p_0/T_0 - p/T) + (N(\mu_0/T_0 - \mu/T) + \ldots$, where $T$ is the temperature, $U$ is the internal energy, $V$ the volume, $p$ the pressure, $N$ the number of particles, and $\mu$ the chemical potential of the system; and the zero index refers to that of the environment (Martinás, Frankowicz 2000).

It is a matter of fact that living systems strive to survive, attempting to avoid death (corresponding to thermal equilibrium) as far and as long as possible. Indeed, life's essential quality has the same unit of measure as action, viz. energy*time. This is because the larger the utilizable energy is for the longer time, the better is the quality and duration of life. Therefore it is apparently inevitable to argue that, in the case of living organisms, the action principle generally obtains not as a minimum, which characterizes physical systems, but as a *maximum*. For living organisms, the length of lifetime is one of the highest priorities. The other is the *ex*tropic energy available for use during a lifetime. Living systems strive not only to mere survival at the edge of death and starvation, but to survive with the maximum of extropic energy. Integrated lifetime extropic energy involves in a single factor both of the highest priorities of life: quality and duration. Living systems' strive to maximum of integrated lifetime extropic energy can be expressed mathematically as corresponding to the maximum version of the action principle.

In a mathematical form, we can formulate the first principle of biology as
$$\int (U(1/T_0 - 1/T) + V(p_0/T_0 - p/T) + N(\mu_0/T_0 - \mu/T) + \ldots)dt = \text{maximum} \qquad (1),$$
where the integral should be taken for the whole lifetime, but which we can use also for the period corresponding to the given physical decay process in which we are interested that would occur in the absence of any biological couplings, such as the coupling of endergonic to exergonic processes. This principle is suitable to determine most biological processes. For example, the dissipation of the internal energy U requires its regeneration through the building up of temperature gradients (first term), mechanical work (second term; lungs, heart etc.), metabolism (third term: μ, N), electromagnetic work (fourth term) etc.

Indeed, this fundamental fact was already recognized by the actual founder of theoretical biology, Ervin Bauer (1967, 53–54): "The living and only the living systems are never in equilibrium, and, on the debit of their free energy, they continuously invest work against the realization of the equilibrium which should occur within the given outer conditions on the basis of the physical and chemical laws" (ibid., 51). He derived all the fundamental phenomena of life such as growth, reproduction, death, regeneration, from this principle, quantitatively. These are significant achievements that indicate we can regard the Bauer principle as the first principle of biology.

We note that the "principle of maximum entropy variation" (Lucia, 2002a,b, 2007) is derived also from the thermodynamic Lagrangian on the basis of the action principle, and has been applied to a mathematical analysis of the dynamics of tumor interaction with the host immune system (Lucia, 2002a), and to the synthesis of ATP molecules (Lucia, 2002b), showing that such general principles may acquire widespread biological applications.

Before going forward, we must consider the problem: on which degrees of freedom does the Bauer principle act?

## Origin of biological possibilities

Biological couplings make otherwise non-spontaneous, endergonic processes possible through coupling them with exergonic processes. In this way, we found that life is the lawful transformation of physically non-spontaneous processes into biologically spontaneous ones, opening up a combinatorial explosion. Therefore, biological problems become definite only if we specify initial and boundary conditions as well as the parameter of optimization. In biological organisms the fundamental coupling occurs not between spatial and temporal coordinates, which would lead to strict constraints on non-thermal degrees of freedom. Rather it occurs with respect to one global or integral thermodynamic state variable: extropy.

Quantum entanglement may also enter the scene. In a certain sense, the living organism can be regarded as one single unified and astronomically complex yet integral gigamolecule, a coherent unit of a whole hierarchy of subsystems that analogously are the megamolecules at subsequent smaller scales, extending down from the organismic level to organs, cells, organelles, supramolecules (Lehn, 1995); and the chemical and physical levels, including again the integral electromagnetic and quantum fields of the whole organism. Therefore, not only the fundamental integral interaction (based on quantum orientation) can play a role in realizing biological needs at the level of the organism, but also the allegedly extraneous capacity for quantum information processing becomes accessible. Definitely, the extraordinary complexity of biological couplings can contribute to the development of higher-level phenomena, such as the organism's self-consciousness.

The presence of highly complex molecules offers easy access for possibilities left open by physics in the form of electronic states of continuous electron bands. These electronic states can carry the fundamental dynamics of biological organization in the presence of significant energy inflows. The result of biological organization into a unified gigamolecule is an enormous set of biologically utilizable and physically not completely determined but still allowed possibilities generated continuously in an astronomical magnitude. All of the astronomical number of biologically generated endergonic possibilities would be extremely improbable in the absence of biological organization. Since autonomous and lawful biological coupling acts between the biologically open possibilities, these otherwise extremely improbable events can develop into a consequent series of biological events, maximizing survival, and optimizing the conditions of life. Evidently animals capable of locomotion can decide almost completely freely which spatial trajectory to select from the enormous set of possibilities. Although the range of such freedom is enormous, all the trajectories should obey the biological constraint to maximize integral extropic energy. Yet the number of remaining biological possibilities is so enormous that different types of optimization parameters are also allowed, such as minimization of the consumed energy, or time till the recovery of original distance from equilibrium.

Biological behavior is flexible. At the same time, there is an invariant aspect of biological behavior, which corresponds to the first principle of biology: investment of work in order to maximize extropic energy on the longest timescale available. Extropy is a global state variable characterizing the system as a whole; therefore, it leaves practically all the other degrees of freedom to be determined by the individual organism.

Actually, integral principles like eq. (1) are so familiar in physics that we have to clarify the step at which the distinguishing biological process, the endpoint selection assume the key role.

### Prototype of biological processes

Let us take as the prototype of all biological processes an extended version of the Galileo experiment, in which we drop a living bird from the Pisa tower. The bird can select different endpoints almost freely, but there are some fundamental types of decisions corresponding to optimization viewpoints. The bird can decide in favor of i) regaining its original height as soon as possible (e.g., in the case where the bird's priority is escaping from the danger zone); ii) regaining its original height with minimal energy investment (the case where the bird's priority corresponds to the case of "no danger" from the external world, but the energy sources are limited), etc. Certainly, some of these fundamental classes of optimization viewpoints are easily calculable. For example, it is easy to determine the time necessary for the bird to regain its nearly original height in case ii) quantitatively. The determination of the trajectory corresponding to the minimal energy investment and the given endpoint is an optimization problem. Wu and Popovic (2003) already construed realistic models of bird flight, in which the trajectory (and the dynamic bird model) are input elements. Definitely, in the absence of the trajectory, such calculations are not possible, since the problem is indefinite. But if the endpoint is already determined by a biological viewpoint, in agreement with the most action principle, the corresponding optimization determines the trajectory. Once the trajectory is obtained, the time development of the bird's flight and the movement of its wings can be determined. Therefore, our method is suitable to enlarge the range of realistic models of the bird flight, and allows adding a further crucial element into the model. With the help of the biological input, the physical model can become suitable to solve the whole biological problem.

Naturally, the bird can fly in any direction, but the corresponding trajectories will have the same form independently of their direction, corresponding to the fact that biology places a constraint only on the distance from equilibrium. And this recognition leads to another, since once the endpoint is (approximately) selected, the problem can be solved by the least action principle, and the fundamental equations of motion are derivable from it.

## Life beyond quantum level

Virtual particles are by definition the ones that do not interact during their lifetime with any other particles. If a virtual particle is born in due time suitable to serve a biological process, and if it is absorbed in due time and due place, activating or triggering biochemically useful reactions, they can be regarded as ideal tools of biological organization simultaneously realizing such fundamental biological functions like timing and biological couplings. If, being virtual, these interactions are instantaneous, this means that biological organization can be much more effective than physical processes. If biological organization harnesses the instantaneous virtual interactions, acting immediately in the whole of the organism, this makes biological organization suitable to realize integral functions of living organisms. Again, the action principle may serve as the physical and biological basis of these integrating processes. Actually, it seems that biological organization is rooted in a level of reality below the quantum level (see also Conrad, Home and Josephson, 1988; Josephson, 2002).

In order to be able to speak about the level of reality beyond the quantum level, we have to outline what we regard as corresponding to the quantum level and what as corresponding to the still deeper, biological level. We think that one obtains the simplest picture if we assume that the virtual interactions of the vacuum can actualize the maximal version of the action principle and generate biological organization in living organisms without any additional factor. Moreover, the virtual interactions can generate the observed physical behavior in all physical systems in which the conditions of biological organization (like high extropy content; large complexity; astronomically large amount of flexible, physically incompletely determined possibilities) are not present. This would mean that the same biological action principle when acting on a physical system can lead naturally to physical behavior.

It seems to be useful to tell a few words about the nature of exergonic-endergonic couplings. Definitely, in living organisms such couplings involve generally many biochemical reactions together. Actually, all biochemical reactions are coupled to each other at the level of cells as well as of the global organism. The living organism is uniquely organized, and no part of it can work independently from the organism without generating malfunctions and illness. Already Ashby (1962, 256) noted that the theory of organization corresponds to non-separable functions. The organism is a coherently organized system of its non-separable biological functions, and these biological functions correspond to an immense number of individual biochemical reactions.

Chauvet (1993, 1995) had shown that functional interactions acting between biological structures are organized hierarchically, leading to the functional organization of the whole organism, resulting to nonlocality corresponding to the basic nature of biological organization according to which all the subsystems of the organism depend on mechanisms that are located elsewhere in the space. The biological organization therefore corresponds always to a closure of functions as well as of biochemical reactions both at the level of the cell and of the organism. This organizational closure, as Matsuno (2006) had shown, can maintain quantum coherence. Recently, McFadden and Al-Khalili (1999) and Davies (2004) considered some processes that can also maintain quantum coherence against physical decoherence processes. Actually, biological organization from time step to time step acts with the help of the action principle, renewing also the functional closure as well as generating the non-computable series of biochemical reactions (Grandpierre, 2007). Since the action principle does not require quantum coherence, biological organization may work also within non-coherent conditions.

This argument is underpinned by the remark that wave functions exist in the Hilbert space. It is a matter of fact that most Hilbert spaces of physics are just multiple manifestations of a single separable Hilbert space. Naturally, biological organization works on the basis of couplings. In living organisms, the different subsystems cannot work independently from each other, and so they cannot be regarded as separable. The activity of each subsystem must be sensitively dependent on the activity of many other subsystems. Therefore all the subsystems of living organisms are coupled to each other; therefore they are not separable by their very biological nature. Therefore, the space corresponding to biological organization must be nonseparable. It is known (Ashby, 1962) that non-metric interactions play a central role in biology and psychology. Since the Hilbert space is metric, therefore biological organization must act at a more general level, in a non-metric space. Again, this requirement is equivalent with our statement that biological organization acts below the quantum level. We note that, plainly, quantum physics is suitable to consider coupled physical systems; but the point is that when the coupling between the physical systems is continuously changing by a law that transcends physical determination, quantum physics becomes insufficient even in principle, independently from the mathematical difficulties.

Therefore a rather surprising situation arises. The two forms of the same action principle — the maximal and the minimal — are *both* used in *living* organisms. For the selection of biological endpoints, living organisms employ the biological action principle (the maximal form). After this first step has been taken, the second step corresponds to optimization, determining the optimal path with the help least action principle. Once the biological endpoint and the parameter of optimization are determined, the biological problem becomes equivalent with fixing the consecutive states of the dropped bird in a way that the corresponding trajectory of the bird leads to the selected endpoint.

In this way, it seems we can enter to a new era of quantitative biology above the molecular level, based on biology meeting physics below the quantum level.

We note that since biology is more fundamental than quantum physics, the bio-friendly nature of the Universe (Davies, 2006) may receive a natural explanation.

## Conclusion

We found a connection between quantum mechanics and the integral character of the action principle which is shown to be suitable to make theoretical biology an ideally mature and quantitative science, indicating that even the integral biological behavior of living organisms can be described by the methods of physics when properly generalized, adjusted, and extended into the realm of biology.

## Acknowledgement

The author wishes to express his gratitude to his friend, Jean Drew, for inspiration, encouragement, for exchanging many exciting ideas, and lecturing the English. This work probably would not have been born without her friendly assistance. It is also a pleasure to thank Dr. Metod Saniga for his careful reading of the draft version of this paper and for valuable suggestions that contributed to significant improvements. It is a pleasure to acknowledge about the useful reference of Engel and Sension advised by Henry Stapp, and about the useful discussions of the physical meaning of action and the action principle with Umberto Lucia.